\documentclass{sig-alternate}

\PassOptionsToPackage{pdfpagelabels=true}{hyperref}
\usepackage{todonotes}
\presetkeys{todonotes}{inline,backgroundcolor=yellow}{}

\usepackage{flushend}

\begin{document}

\newcommand{\ttv}{\textit{Text2Vis}}

\setcopyright{rightsretained}

\doi{}

\isbn{}

\conferenceinfo{Neu-IR '16 SIGIR Workshop on Neural Information Retrieval}{July 21, 2016, Pisa, Italy}

\copyrightetc{\noindent\confname\ \the\conf\ \the\confinfo\par\smallskip  \copyright\ {\normalfont 2016 Copyright held by the owner/author(s).}}

\title{Picture It In Your Mind: Generating High Level Visual Representations From Textual Descriptions}

\numberofauthors{5} 
\author{
\alignauthor
Fabio Carrara\\
       \affaddr{ISTI-CNR}\\
       \affaddr{via G. Moruzzi, 1}\\
       \affaddr{56124 Pisa, Italy}\\
       \email{fabio.carrara@isti.cnr.it}
\alignauthor
Andrea Esuli\\
       \affaddr{ISTI-CNR}\\
       \affaddr{via G. Moruzzi, 1}\\
       \affaddr{56124 Pisa, Italy}\\
       \email{andrea.esuli@isti.cnr.it}
\alignauthor
Tiziano Fagni\\
       \affaddr{ISTI-CNR}\\
       \affaddr{via G. Moruzzi, 1}\\
       \affaddr{56124 Pisa, Italy}\\
       \email{tiziano.fagni@isti.cnr.it}
\and 
\alignauthor
Fabrizio Falchi\\
       \affaddr{ISTI-CNR}\\
       \affaddr{via G. Moruzzi, 1}\\
       \affaddr{56124 Pisa, Italy}\\
       \email{fabrizio.falchi@isti.cnr.it}
\alignauthor
Alejandro Moreo Fern\'andez\\
       \affaddr{ISTI-CNR}\\
       \affaddr{via G. Moruzzi, 1}\\
       \affaddr{56124 Pisa, Italy}\\
       \email{alejandro.moreo@isti.cnr.it}
}

\maketitle
\begin{abstract}
In this paper we tackle the problem of image search when the query is a short textual description of the image the user is looking for.
We choose to implement the actual search process as a similarity search in a visual feature space, by learning to translate a textual query into a visual representation.
Searching in the visual feature space has the advantage that any update to the translation model does not require to reprocess the, typically huge, image collection on which the search is performed.
We propose \ttv{}, a neural network that generates a visual representation, in the visual feature space of the fc6-fc7 layers of ImageNet, from a short descriptive text.
\ttv{} optimizes two loss functions, using a stochastic loss-selection method.
A visual-focused loss is aimed at learning the actual text-to-visual feature mapping, while a text-focused loss is aimed at modeling the higher-level semantic concepts expressed in language and countering the overfit on non-relevant visual components of the visual loss.
We report preliminary results on the MS-COCO dataset.
\end{abstract}

\printccsdesc


\section{Introduction}
\label{sec:intro}

Using a textual query to retrieve images is a very common cross-media search task, as text is the most efficient media to describe the kind of image the user is searching for.
The actual retrieval process can be implemented in a number of ways, depending on how the shared search space between text and images is defined.
The search space can be based on textual features, visual features, or a joint space in which textual and visual features are projected into.

Using textual features is the most common solution, specially at the Web scale.
Each image is associated with a set of textual features extracted from its context of use (e.g., the text surrounding the image in the Web page, description fields in metadata), and eventually enriched by means of classifiers that assign textual labels related to the presence or certain relevant entities or abstract properties in the image.
The textual search space model can exploit the actual visual content of the image only when classifiers for the concepts of interest are available, thus requiring a relevant number of classifiers; this also requires to reprocess the entire image collection whenever a new classifier is made available.

On the other side, the visual and joint search spaces represent each image through visual features extracted from its actual content.
The method we propose in this paper adopts a visual space search model.
A textual query is converted into a visual representation in a visual space, where the search is performed by similarity.
An advantage of this model is that any improvement in the text representation model, and its conversion to visual features, has immediate benefits on the image retrieval process, without requiring to reprocess the whole image collection.

A joint space model requires instead a reprocessing of all images whenever the textual model is updated, since the projection of images into the joint space is influenced also by the textual model part.
It also requires managing and storing the additional joint space representations that are used only for the cross-media search.

In this paper we present the preliminary results on learning \ttv{}, a neural network model that converts textual descriptions into visual representations in the same space of those extracted from deep Convolutional Neural Networks (CNN) such as ImageNet \cite{krizhevsky2012imagenet}.
\ttv{} achieves its goal by using a stochastic loss choice on two separate loss functions (as detailed in Section \ref{sec:method}), one for textual representations autoencoding, and one for visual representations generation.
Preliminary results show that the produced visual representations capture the high level concepts expressed in the textual description.

\section{Related work}

Deep Learning and Deep Convolutional Neural Networks (DCNNs) in particular, have recently shown impressive performance on a number of multimedia information retrieval tasks \cite{krizhevsky2012imagenet,simonyan2014very,he2015deep}.
Deep Learning methods learn representations of data with multiple levels of abstraction.
As a result, the activation of the hidden layers has been used in the context of transfer learning and content-based image retrieval \cite{donahue2013decaf,razavian2014cnn} as high-level representations of the visual content.
Somewhat similarly, distributional semantic models, such as those produced by Word2Vec \cite{mikolov2013distributed}, or GloVe \cite{pennington2014glove}, have been found useful in modeling semantic similarities among words by establishing a correlation between \emph{word meaning} and \emph{position} in a vector space.

In order to perform cross-media retrieval, the two feature spaces (text and images in our case) should become comparable, typically by learning how to properly map the different sources. 
This problem has been attempted in different manners so far, which could be roughly grouped into three main variants, depending on whether the mapping is performed into a common space, the textual space, or the visual space.

\textbf{Mapping into a common space:}
The idea of comparing texts and images in a shared space has been investigated by means of Cross-modal Factor Analysis and (Kernel) Canonical Correlation Analysis in \cite{costa2014role}.
In a similar vein, Corr-AE was proposed for cross-modal retrieval, allowing the search to be performed in both directions, i.e., from text-to-image and viceversa \cite{feng2014cross}.
The idea is to train two autoencoders, one for the image domain and another for the textual domain, imposing restrictions between the two.
As will be seen, the architecture we are presenting here bears resemblance to one of the architectures investigated in \cite{feng2014cross}, the so-called \emph{Correspondence full-modal autoencoder} (which is inspired by the multimodal deep learning method \cite{ngiam2011multimodal}).
Contrarily to the multimodal architectures though, we apply a stochastic criterion to jointly optimize for the two modals, thus refraining from combining them into a parametric single loss.

\textbf{Mapping into the textual space:}
The BoWDNN method trains a deep neural network (DNN) to map images directly into a bag-of-words (BoW) space, where the cosine similarity between BoWs representations is used to generate the ranking \cite{bai2014bag}.
Somehow similarly, a dedicated area of related research is focused on generating captions describing the salient information of an image (see, e.g., \cite{karpathy2015deep, fang2015captions}).

Two other important examples along these lines are DeViSE \cite{frome2013devise} and ConSE \cite{norouzi2013zero}. 
Both methods build upon the higher layers of the convolutional neural network of \cite{krizhevsky2012imagenet}; the main difference lies on the way both methods treat the last layer of the net. 
Whereas DeViSE replaces this last layer with a linear mapping (thus fine-tuning the whole network) ConSE, on the other side, directly takes the outputs of the last layer and learns a projection to the textual embedding space.

\textbf{Mapping into the visual space:}
Our proposal \ttv{} belongs to this group where, to the best of our knowledge, the only example up to now was a method dubbed \emph{Word2VisualVec} \cite{Dong2016}, which was reported just very recently.
There are some fundamental points where their method and ours differ, though.
On the one hand, their \emph{Word2VisualVec} takes combinations of Word2Vec-like vectors as a starting point, thus reducing the dimensionality of the input space; we directly start from the bag-of-words vector encoding of the textual space, as we did not observed any improvement in pre-training the textual part. 
On the other, they build a deep network on top of the textual representation.
As shall be seen, our \ttv{} is much shallower, as we found the net to be capable of mapping textual vectors into the visual space quite efficiently, provided that the model is properly regularized; an issue on which we focused our attention.

\section{Generating visual representations of text}
\label{sec:method}

\noindent In this section we describe the architecture of our \ttv{} network. 
Our idea is to map textual descriptions to high-level visual representations.
As the visual space we used the \emph{fc6} and \emph{fc7} layers of the Hybrid network \cite{zhou2014learning} (i.e., an AlexNet \cite{krizhevsky2012imagenet} trained on both ImageNet\footnote{\url{http://image-net.org}} and Places\footnote{\url{http://places.csail.mit.edu/index.html}} datasets).
We tested two vectorial representations for the textual descriptions: \ttv{}$_1$ uses simple bag-of-words vectors that mark with a value of one the positions that are relative to words that appear in the textual description and leave to zero all the others; \ttv{}$_N$ adds a bit text structure info by considering also N-grams for a selection of part-of-speech patterns\footnote{We considered the part-of-speech patterns: `NOUN-VERB', `NOUN-VERB-VERB', `ADJ-NOUN', `VERB-PRT', `VERB-VERB', `NUM-NOUN', and `NOUN-NOUN'.}.
\ttv{}$_N$ is a first approach at modeling text structure into the input vectorial representation, which differentiates the task of search from detailed/complex textual description we aim at from the traditional keyword search.
 
We have also investigated the use of pre-trained word embeddings, representing the textual description as the average of the embeddings of the words composing the description (see Equation 1 in \cite{Dong2016}), but we have not observed any improvement.
Generating the word embeddings is an additional cost, and the fitness of the embeddings for the task depends on the type of documents they are learned from.
For example, an 11\% improvement in MAP is reported in \cite{Cappallo2015} from learning embedding from Flickr tags compared to learning them from Wikipedia pages.
The direct use of bag-of-words vectors in \ttv{} removes the variable of selecting an appropriate document collection to learn the embedding and its learning cost.

\begin{figure*}[t]
 \includegraphics[width=\textwidth]{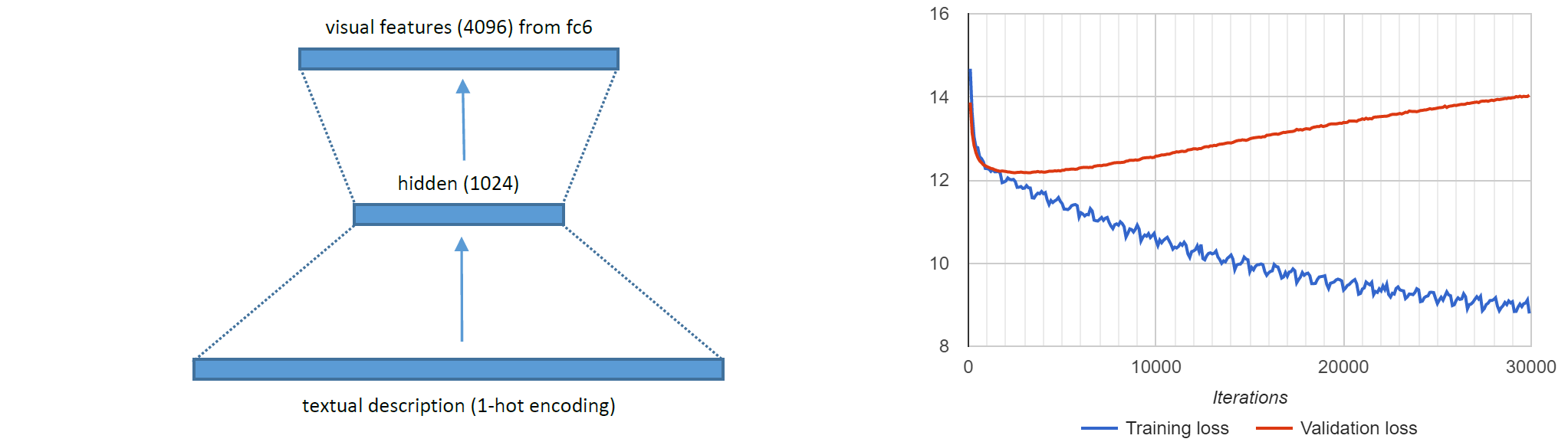}
 \caption{Overfitting of a simple regressor model with one hidden layer of size 1024.}
 \label{fig:reg:overfit}
 \end{figure*}

As described in the following, \ttv{} actually learns a description embedding space that is able to reconstruct both the original description and the visual description.
To reach this, we started with a simple regressor model (Figure \ref{fig:reg:overfit}, left) trained to directly predict the visual representation of the image associated with the textual input.
We observed a strong tendency to overfit (Figure \ref{fig:reg:overfit}, right), thus degrading the applicability of the method to unseen images. 

We explained this overfitting with the fact that a visual representation keeps track of every element that appears in the image, regardless of their semantic relevance within the image, while a (short) textual description is more likely focused on the visually relevant information, disregarding the secondary content of the image, as shown in Figure \ref{sec:cases}.
As the learning iterations proceed, the simple regressor model starts capturing secondary elements of the images that are not relevant for the main represented concept, but are somewhat characteristic in the training data.

\begin{figure*}[t]
\includegraphics[width=\textwidth]{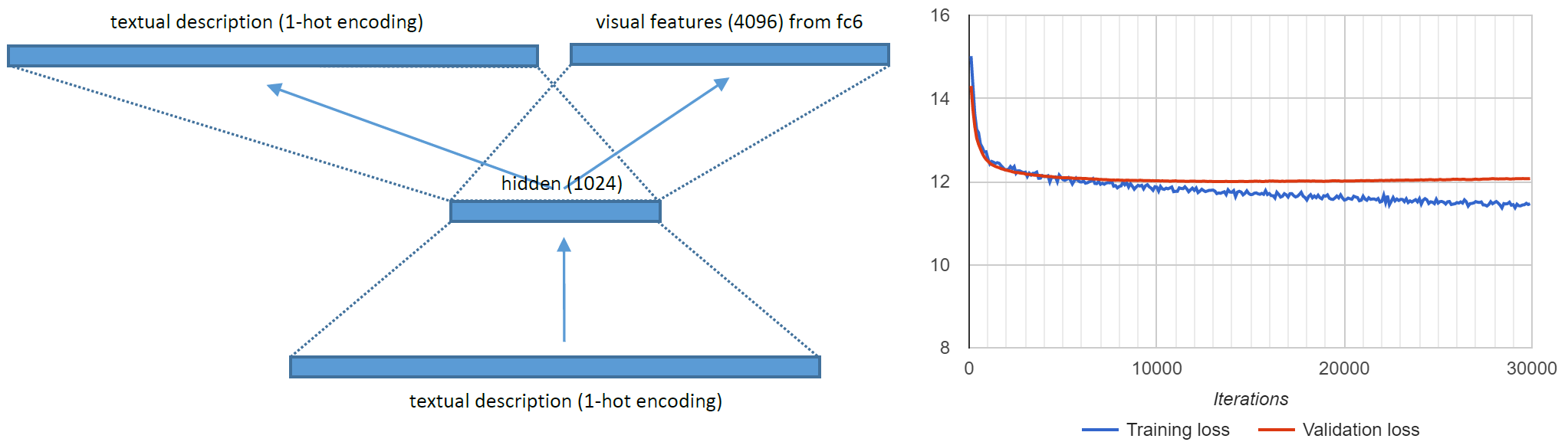}
\caption{Our proposed \ttv{} which controls overfitting by adding an autoencoding constraint on the hidden state.}
\label{fig:net-scheme}
\end{figure*}

Our \ttv{} proposal to contrast such overfitting is to add a text-to-text autoencoding branch to the hidden layer (Figure \ref{fig:net-scheme}, left), forcing the model to satisfy two losses: one visual (text-to-visual regression) and one linguistic (text-to-text autoencoder).
The linguistic loss works at higher level of abstraction than the visual one, acting as a regularization constraint on the model, and preventing, as confirmed by our experiments, overfitting on the visual loss (Figure \ref{fig:net-scheme}, right).
As detailed in the next section, we implemented the use of the two losses with a stochastic process, in which at each iteration one of the two is selected for optimization.

\subsection{\ttv{}}\label{subsec:ttv}

\ttv{} consists of two overlapped feedforward neural nets with a shared hidden layer.
The shared hidden layer causes a regularization effect during the combined optimization; i.e., the hidden state is constrained to be a good representation to accomplish with two different goals.
The feedforward computation is described by the following equations:
\begin{eqnarray}
z = ReLU(W_1 t_{in} + b_1) \label{eqn:tin-z}\\
t' = ReLU(W_2 z + b_2) \label{eqn:z-tout}\\
v' = ReLU(W_3 z + b_3) \label{eqn:z-vout}
\end{eqnarray}
\noindent where $t_{in}$ represents the bag-of-words encoding for the textual descriptor given as input to the net, $z$ is the hidden representation, $v'$ and $t'$ are the visual and textual predictions, respectively, obtained from the hidden representation $z$, $\Theta=\{W_i,b_i\}_{i\in\{1,2,3\}}$ are the model parameters to be learned, and $ReLU$ is the  activation function, defined by $ReLU(x)=\max\{0,x\}$.

Both predictions $v'$ and $t'$ are then confronted with the expected outputs (i) the visual representation $v$ corresponding to the $fc6$ or $fc7$ layers of \cite{krizhevsky2012imagenet}, and (ii) a textual descriptor $t_{out}$ that is semantically equivalent to $t_{in}$.
We used the \emph{mean squared error} (MSE) as the loss function in both cases:
\begin{equation}
\mathcal{L}(x,y;\Theta') = MSE(x,y) = \frac{1}{n}\sum_{i=1}^{n}(x_i-y_i)^2
\end{equation}
The model is thus multi-objective, and many alternative strategies could be followed at this point in order to set the $\Theta$ parameters so that both criteria are jointly minimized.
We rather propose a much simpler, yet effective, way for carrying out the optimization search, that consists of considering both branches of the net as independent, and randomly deciding in each iteration which of them is to be used for the gradient descend optimization.

Let thus define $\Theta_t=\{W_i,b_i\}_{i\in\{1,2\}}$ and $\Theta_v=\{W_i,b_i\}_{i\in\{1,3\}}$ as the model parameters of each independent branch. 
The optimization problem has two objectives (Equations \ref{eq:op:t} and \ref{eq:op:v}), and at each iteration, a random choice decides which of them is to be optimized.  
We call this heuristic the \emph{Stochastic Loss} (SL) optimization.
\begin{eqnarray}
 \Theta_{\widehat{t}}  =  argmin_{\Theta_t} \mathcal{L}_t(t_{out},t';\Theta_t) \label{eq:op:t}\\
 \Theta_{\widehat{v}}  =  argmin_{\Theta_v} \mathcal{L}_v(v,v';\Theta_v) \label{eq:op:v}
\end{eqnarray}

Note that the net is fed with a triple $\langle v,t_{in},t_{out} \rangle$ at each iteration.
When $t_{out}=t_{in}$ the text-to-text branch is an \emph{autoencoder}.
It is also possible to have $t_{in}\neq t_{out}$, with the two pieces of text been semantically equivalent (e.g., $t_{in}=$``\texttt{a woman cutting a pizza with a knife}'', $t_{out}=$``\texttt{a woman holds a knife to cut pizza}'') then the text-to-text branch might be reminiscent of the \emph{Skip-gram}- and \emph{CBOW}-like architectures.
The text-to-image branch is, in any case, a regressor.
The SL causes the model to be co-regularized. 
Notwithstanding, since our final goal is to project the textual descriptor into the visual space, the text-to-text branch might be though as a regularization to the visual reconstruction (and, more specifically, to its internal encoding) which responds to constrains of linguistic nature.

\section{Experiments}
\subsection{Datasets}

We used the \emph{Microsoft COCO} dataset (MsCOCO\footnote{Publicly available at \url{http://mscoco.org/}} \cite{lin2014microsoft}). 
MsCOCO was originally proposed for image recognition, segmentation, and caption generation. 
Although other datasets for image retrieval exist (e.g., the one proposed in \cite{hua2013clickage}), they are more oriented to keyword-based queries.
We believe MsCOCO to be more fit to the scenario we want to explore, since the captions associated to the images are expressed in natural language, thus semantically richer than a short list of keywords composing a query.

MsCOCO contains 82.783 training images (\emph{Train2014}), 40.504 validation images (\emph{Val2014}), and about 40K and 80K test images corresponding to two different competitions \cite{chen2015microsoft} (\emph{Test2014} and \emph{Test2015}).
Because MsCOCO was proposed for caption generation, the captions are only accessible in the \emph{Train2014} and \emph{Val2014} sets, while they are not yet released for \emph{Test2014} and \emph{Test2015}.
We have thus taken the \emph{Train2014} set for training, and split the \emph{Val2014} into two disjoint sets of 20K images each for validation and test.

Each image in MsCOCO has 5 different captions associated.
Let $\langle I,C\rangle$ be any labeled instance in MsCOCO, where $I$ is an image and $C=\{c_1..c_5\}$ is a set of captions describing the content of $I$.
Given a $\langle I,C\rangle$ pair, we define a labeled instance in our model as $\langle v,t_{in},t_{out}\rangle$, where $v\in\mathbb{R}^{4096}$ is the visual representation of the image $I$ taken from the \emph{fc6} layer (or \emph{fc7}, in separate experiments) of the Hybrid network \cite{zhou2014learning}; $t_{in}$ and $t_{out}$ are two textual descriptors from $C$ representing the input and output descriptors for the model, respectively.
During training, $t_{in}$ and $t_{out}$ are uniformly chosen at random from $C$ (thus $t_{in}$ and $t_{out}$ are not imposed to be different).
Note that the number of training instances one could extract from a given $\langle I,C\rangle$ amounts to 25, which increases the variability of the training set along the different epochs.


\subsection{Training}

We solve the optimization problems of Equations \ref{eq:op:t} and \ref{eq:op:v}, using the \emph{Adam} method \cite{adam} for stochastic optimization, with default parameters (learning rate $\alpha=0.001$, $\beta_1=0.9$, $\beta_2=0.999$, and $\epsilon=1e^{-0.8}$). 
Note that there are two independent instances of the Adam optimizer, one associated to $\mathcal{L}_t$ (Equation \ref{eq:op:t}) and other for $\mathcal{L}_v$ (Equation \ref{eq:op:v}).
In this preliminary study we decided to set for the SL an equal selection probability to both $\mathcal{L}_t$ and $\mathcal{L}_v$; different distributions will be investigated in future research.

We set the size of the training batch to 100 examples.
We set the maximum number of iterations to 300.000, but apply an early stop when the model starts overfitting (as reflected in the validation error).
The training set is shuffled each time a complete pass over all images is completed.

All the $\Theta$ parameters have been initialized at random according to a truncated normal distribution centered in zero with standard deviation of $\frac{1}{\sqrt{n}}$, where $n$ is the number of columns. The biases have all been initialized to 0. 

The vocabulary size is 10,358 for \ttv{}$_1$ after removing terms appearing in less than 5 captions.
For \ttv{}$_N$ we considered the 23,968 uni-grams and N-grams appearing at least in 10 captions.
Since the number of units in the hidden and output layers are 1024 and 4096, respectively, the total number of parameters of the models amount to 
25.4M 
in \ttv{}$_1$ and 
53.3M 
in \ttv{}$_N$.

A Tensorflow implementation of \ttv{} is available at \url{https://github.com/AlexMoreo/tensorflow-Tex2Vis}.
\begin{figure}[t]
\begin{center}
\includegraphics[width=\columnwidth]{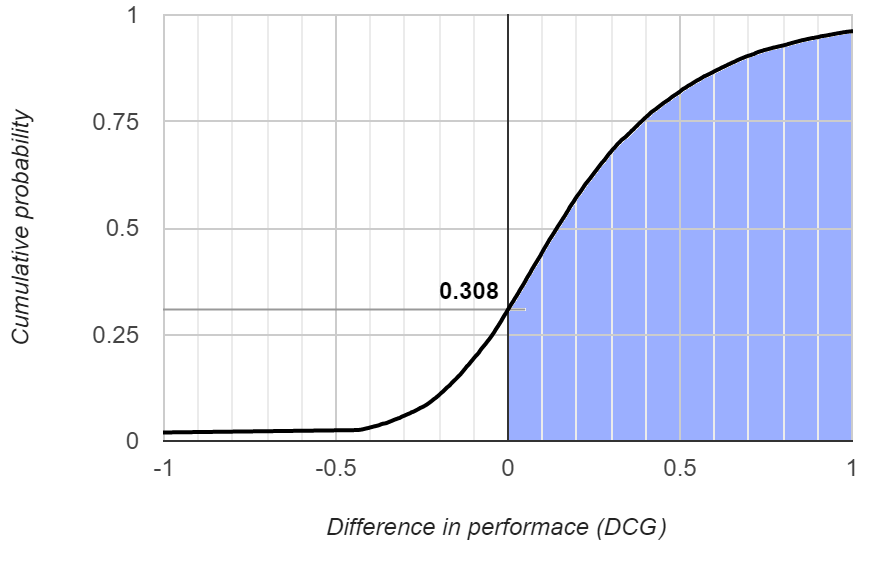}\vspace{-4ex}
\includegraphics[width=\columnwidth]{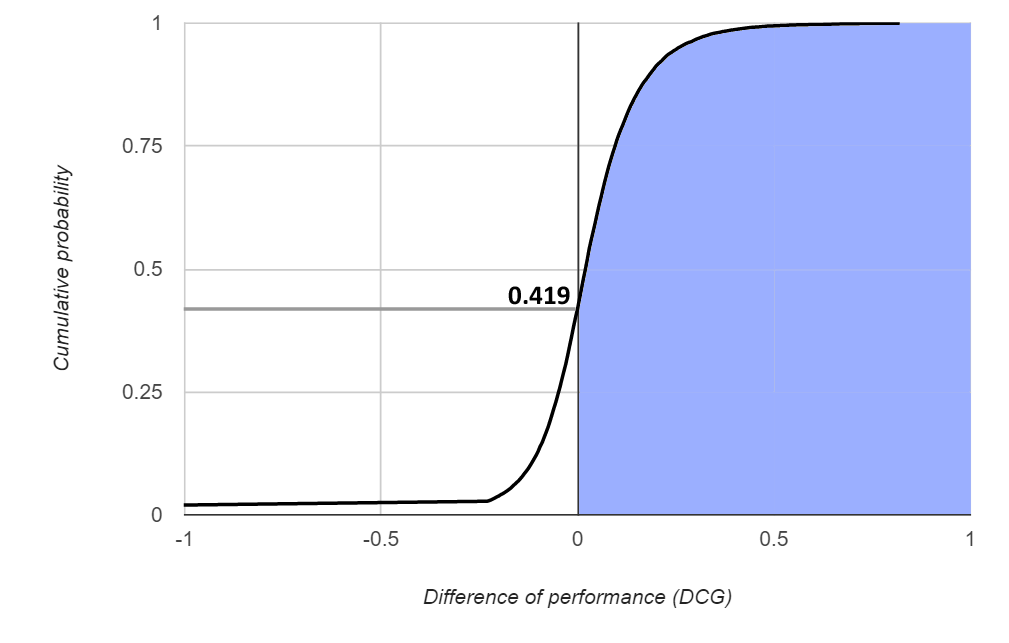}
\end{center}
\caption{Cumulative probability distribution of the difference in performance of our \ttv{}$_1$ with respect to \emph{VisSim} (upper plot) and \emph{VisReg} (lower plot), on $fc6$.
Positive differences mean \ttv{} obtained a better ranking score than \emph{VisSim} or \emph{VisReg} (resp. 69.2\% and 58.1\% of cases, shadowed region).
}
\vspace{-2ex}\label{fig:dcgr_dist}
\end{figure}

\subsection{Evaluation Measures}

Image retrieval is performed by similarity search in the visual space, using Euclidean distance on the l2-normalized visual vectors to generate a ranking of images, sorted by closeness. 
We measure the retrieval effectiveness of the visual representations produced from textual descriptions by our \ttv{} network by means of the \emph{Discounted Cumulative Gain} (DCG \cite{dcg}), defined as:
\begin{equation}
DCG_p=\sum_{i=1}^{p}\frac{2^{rel_i}-1}{\log_2(i+1)}
\end{equation}
\noindent where $rel_i$ quantifies the \emph{relevance} of the retrieved element at rank position $i$ with respect to the query, and $p$ is the rank at which the metric is computed; we set $p=25$ in our experiments, as was done in related research \cite{hua2013clickage,Dong2016}.

Because the $rel$ values are not provided in the MsCOCO, we estimate them by using the $ROUGE_{L}$ \cite{ROUGE} metric.
$ROUGE_{L}$ is one of the evaluation measures for the MsCOCO caption generation competition\footnote{\url{https://github.com/tylin/coco-caption}} \cite{chen2015microsoft}.
We compute $rel_i=ROUGE_{L}(t_{in}, C_i)$, where $t_{in}$ is the query caption, and $C_i$ are the 5 captions associated to the retrieved image at rank $i$.
This caption-to-caption relevance model is thus aimed at measuring how much the concepts expressed in the query appear as relevant parts of the retrieved images.

\subsection{Results}

We compared the performance of \ttv{}$_1$ and \ttv{}$_N$ models against:
\emph{RRank}, a lower bound baseline that produces a random ranking of images, for any query;
\emph{VisSim}, a direct similarity method that computes the Euclidean distances using the original \emph{fc6}, or \emph{fc7}, features for the image that is associated to query caption in MsCOCO; 
and \emph{VisReg}, the text-to-image regressor described in Figure \ref{fig:reg:overfit}.

Table \ref{tab:dcg} reports the averaged DCG scores obtained by the compared methods.
These results show a significant improvement of our proposal with respect to the compared methods.
When using $fc6$ as the visual space, \ttv{}$_1$ obtains a $8.51\%$ relative improvement with respect to \emph{VisSim} and $1.40\%$ over \emph{VisReg}. 
The improvements of \ttv{}$_N$ are respectively of $8.08\%$ and $0.94\%$.
When using $fc7$ as the visual space it is \ttv{}$_N$ that obtains, yet by a small margin, the best result. The relative improvements of \ttv{}$_1$ over emph{VisSim} and \emph{VisReg} are respectively of $8.48\%$ and $0.97\%$, and for \ttv{}$_N$ respectively of $8.60\%$ and $1.09\%$.

\begin{table}[ht!]
\center
\begin{tabular}{|l|c|c|}
\hline
Method 	& fc6 & fc7 \\ \hline
\emph{RRank}	& 1.524 & 1.524 \\
\emph{VisSim}	& 2.150 & 2.180 \\
\emph{VisReg}	& 2.317 & 2.359 \\
\ttv{}$_{1}$	& \textbf{2.350} & 2.382 \\
\ttv{}$_{N}$	& 2.339 & \textbf{2.385} \\\hline
\end{tabular}
\label{tab:dcg}
\caption{Performance comparison of the different methods in terms of averaged DCG}
\end{table}

In addition to the averaged performance, we also investigated how often the ranking produced by \ttv{} is more relevant (according to DCG) than those produced by \emph{VisSim} and \emph{VisReg}. 
Figure \ref{fig:dcgr_dist} indicates that in 69.2\% of the cases, the ranking of \ttv{}$_1$ was found more relevant than \emph{VisSim} (see Figure \ref{fig:dcgr_dist}).
The same happens in $58.1\%$ of the cases when comparing \ttv{} to \emph{VisReg}.

\begin{figure}[!b]
\center
\includegraphics[width=\columnwidth]{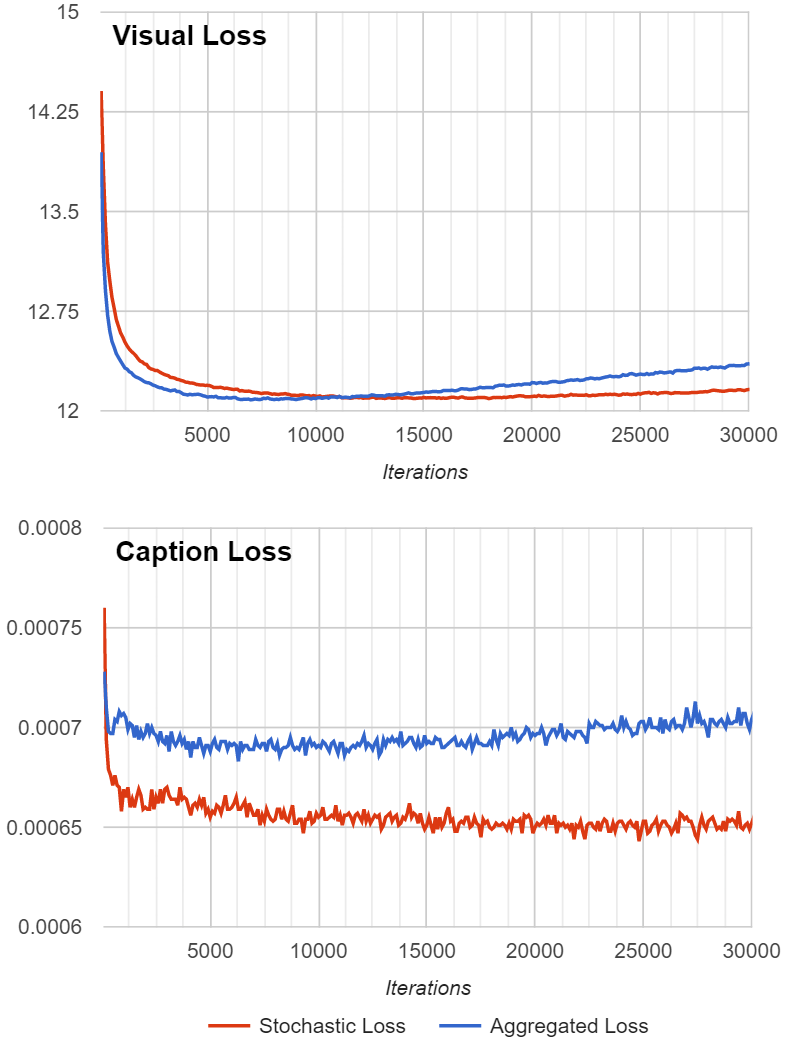}
\caption{Validation loss for $\mathcal{L}_v$ and $\mathcal{L}_t$, optimizing on a linear combination of losses (blue) or using two  optimizers with stochastic loss selection (SL, red).}
\label{fig:modaltrends}
\end{figure}

\begin{figure*}[ht!]
\begin{center}
\includegraphics[width=0.88\textwidth]{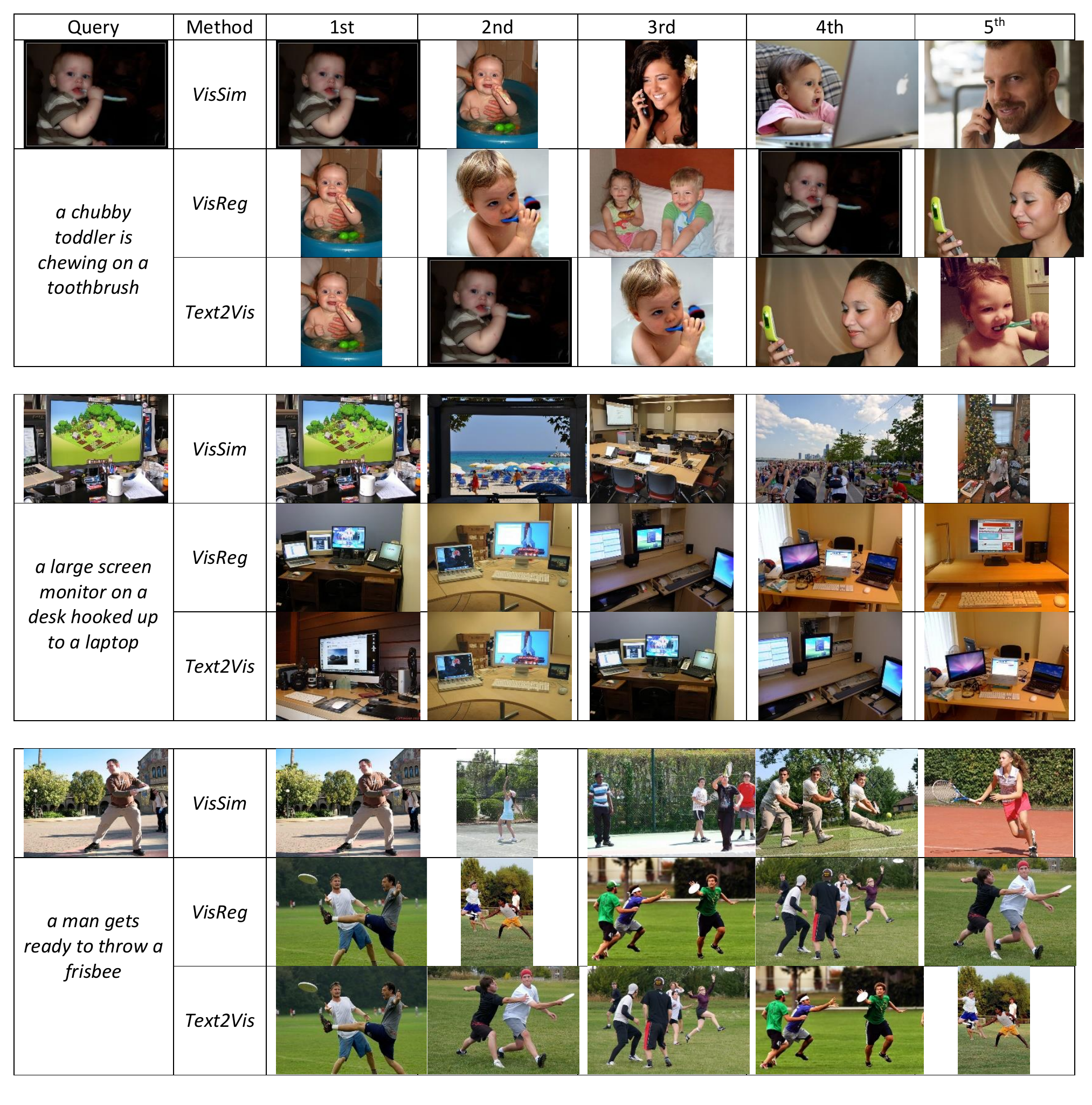}
\end{center}
\caption{Examples of search results from the three compared methods.}
\label{fig:viscomp}
\end{figure*}
\subsection{Why Stochastic Loss?}\label{sec:sl}
\ttv{} uses two independent optimizers to optimize the visual ($\mathcal{L}_v$) and the textual ($\mathcal{L}_t$) losses, based on a stochastic choice at each iteration (SL, section \ref{subsec:ttv}). 
Previous approaches to multimodal learning relied instead on a unique aggregated loss (typically of the form $\mathcal{L}=\mathcal{L}_v+\lambda\mathcal{L}_t$) that is minimized by a single optimizer \cite{feng2014cross,ngiam2011multimodal}.
We compared the two approaches on the case of equal relevance of the two losses ($\lambda=1$, uniform distribution for SL).
SL better optimizes the two losses (Figure \ref{fig:modaltrends}), and is less prone to overfit.

We deem that SL allows to model in a more natural way the relative relevance of the various losses that are combined, i.e., by selecting the losses in proportion to the assigned relevance, whereas the numeric aggregation is affected by the relative values of losses and the differences in their variation during the optimization (e.g., a loss that has a large improvement may compensate for another loss getting worse).
SL is also computationally lighter than the aggregated loss, as SL updates only a part of the model on each iteration.

\subsection{Visual comparison}\label{sec:cases}

Figure \ref{fig:viscomp} show a few samples\footnote{More results at \url{https://github.com/AlexMoreo/tensorflow-Tex2Vis}} that highlight the differences in results from the three compared methods.
In all the cases results from the \emph{VisSim} method are dominated by the main visual features of the images: a face for the first query, the content of the screen for the second query, an outdoor image with a light lower part, plants, people and a bit of sky in the third one.
The two text based methods obtains results that more often contain the key elements of the description. 
For the first query, \ttv{} retrieves four relevant images out of five, one more that \emph{VisReg}.
For the other two queries the results are pretty similar, with \ttv{} placing in second position an image that is a perfect match for the query, while \emph{VisReg} places it in fifth position.

\section{Conclusions}
The preliminary experiments indicate our method produces more relevant rankings than those produced by similarity search directly on the visual features of a query image.
This is an indication that our text-to-image mapping produces better prototypical representations of the desired scene than the representation of a sample image itself.
A simple explanation of this result is that textual descriptions strictly emphasize the relevant aspects of the scene the user has in mind, whereas the visual features, directly extracted from the query image, are keeping track of all the information that is contained in that image, causing the similarity search to be potentially confused by secondary elements of the scene.
The \ttv{} model also improved, yet by a smaller margin, over the \emph{VisReg} model , showing that an auto-enconding branch in the network is useful to avoid overfitting on visual features.
We also found that combing losses in a stochastic fashion, rather than numerically, improves both the effectiveness and efficiency of the system. 
In the future we plan to compare \ttv{} against the recently proposed \emph{Word2VisualVec} \cite{Dong2016} model.
We also intend to improve the modeling word order information in \ttv{}, likely by adding a recurrent component to the network architecture.

\bibliographystyle{abbrv}
\bibliography{sigproc} 

\begin{thebibliography}{10}

\bibitem{bai2014bag}
Y.~Bai, W.~Yu, T.~Xiao, C.~Xu, K.~Yang, W.-Y. Ma, and T.~Zhao.
\newblock Bag-of-words based deep neural network for image retrieval.
\newblock In {\em Proceedings of the ACM International Conference on
  Multimedia}, pages 229--232. ACM, 2014.

\bibitem{Cappallo2015}
S.~Cappallo, T.~Mensink, and C.~G. Snoek.
\newblock Image2emoji: Zero-shot emoji prediction for visual media.
\newblock In {\em Proceedings of the 23rd ACM International Conference on
  Multimedia}, MM '15, pages 1311--1314, New York, NY, USA, 2015. ACM.

\bibitem{chen2015microsoft}
X.~Chen, H.~Fang, T.-Y. Lin, R.~Vedantam, S.~Gupta, P.~Doll{\'a}r, and C.~L.
  Zitnick.
\newblock Microsoft coco captions: Data collection and evaluation server.
\newblock {\em arXiv preprint arXiv:1504.00325}, 2015.

\bibitem{costa2014role}
J.~Costa~Pereira, E.~Coviello, G.~Doyle, N.~Rasiwasia, G.~R. Lanckriet,
  R.~Levy, and N.~Vasconcelos.
\newblock On the role of correlation and abstraction in cross-modal multimedia
  retrieval.
\newblock {\em Pattern Analysis and Machine Intelligence, IEEE Transactions
  on}, 36(3):521--535, 2014.

\bibitem{donahue2013decaf}
J.~Donahue, Y.~Jia, O.~Vinyals, J.~Hoffman, N.~Zhang, E.~Tzeng, and T.~Darrell.
\newblock Decaf: A deep convolutional activation feature for generic visual
  recognition.
\newblock {\em arXiv preprint arXiv:1310.1531}, 2013.

\bibitem{Dong2016}
J.~{Dong}, X.~{Li}, and C.~G.~M. {Snoek}.
\newblock {Word2VisualVec: Cross-Media Retrieval by Visual Feature Prediction}.
\newblock {\em ArXiv e-prints}, Apr. 2016.

\bibitem{fang2015captions}
H.~Fang, S.~Gupta, F.~Iandola, R.~K. Srivastava, L.~Deng, P.~Doll{\'a}r,
  J.~Gao, X.~He, M.~Mitchell, J.~C. Platt, et~al.
\newblock From captions to visual concepts and back.
\newblock In {\em Proceedings of the IEEE Conference on Computer Vision and
  Pattern Recognition}, pages 1473--1482, 2015.

\bibitem{feng2014cross}
F.~Feng, X.~Wang, and R.~Li.
\newblock Cross-modal retrieval with correspondence autoencoder.
\newblock In {\em Proceedings of the ACM International Conference on
  Multimedia}, pages 7--16. ACM, 2014.

\bibitem{frome2013devise}
A.~Frome, G.~S. Corrado, J.~Shlens, S.~Bengio, J.~Dean, T.~Mikolov, et~al.
\newblock Devise: A deep visual-semantic embedding model.
\newblock In {\em Advances in Neural Information Processing Systems}, pages
  2121--2129, 2013.

\bibitem{he2015deep}
K.~He, X.~Zhang, S.~Ren, and J.~Sun.
\newblock Deep residual learning for image recognition.
\newblock {\em arXiv preprint arXiv:1512.03385}, 2015.

\bibitem{hua2013clickage}
X.-S. Hua, L.~Yang, J.~Wang, J.~Wang, M.~Ye, K.~Wang, Y.~Rui, and J.~Li.
\newblock Clickage: Towards bridging semantic and intent gaps via mining click
  logs of search engines.
\newblock In {\em Proceedings of the 21st ACM international conference on
  Multimedia}, pages 243--252. ACM, 2013.

\bibitem{dcg}
K.~J{\"a}rvelin and J.~Kek{\"a}l{\"a}inen.
\newblock Cumulated gain-based evaluation of ir techniques.
\newblock {\em ACM Transactions on Information Systems (TOIS)}, 20(4):422--446,
  2002.

\bibitem{karpathy2015deep}
A.~Karpathy and L.~Fei-Fei.
\newblock Deep visual-semantic alignments for generating image descriptions.
\newblock In {\em Proceedings of the IEEE Conference on Computer Vision and
  Pattern Recognition}, pages 3128--3137, 2015.

\bibitem{adam}
D.~Kingma and J.~Ba.
\newblock Adam: A method for stochastic optimization.
\newblock {\em arXiv preprint arXiv:1412.6980}, 2014.

\bibitem{krizhevsky2012imagenet}
A.~Krizhevsky, I.~Sutskever, and G.~E. Hinton.
\newblock Imagenet classification with deep convolutional neural networks.
\newblock In {\em Advances in neural information processing systems}, pages
  1097--1105, 2012.

\bibitem{ROUGE}
C.-Y. Lin.
\newblock Rouge: A package for automatic evaluation of summaries.
\newblock In S.~S. Marie-Francine~Moens, editor, {\em Text Summarization
  Branches Out: Proceedings of the ACL-04 Workshop}, pages 74--81, Barcelona,
  Spain, July 2004. Association for Computational Linguistics.

\bibitem{lin2014microsoft}
T.-Y. Lin, M.~Maire, S.~Belongie, J.~Hays, P.~Perona, D.~Ramanan,
  P.~Doll{\'a}r, and C.~L. Zitnick.
\newblock Microsoft coco: Common objects in context.
\newblock In {\em Computer Vision--ECCV 2014}, pages 740--755. Springer, 2014.

\bibitem{mikolov2013distributed}
T.~Mikolov, I.~Sutskever, K.~Chen, G.~S. Corrado, and J.~Dean.
\newblock Distributed representations of words and phrases and their
  compositionality.
\newblock In {\em Advances in neural information processing systems}, pages
  3111--3119, 2013.

\bibitem{ngiam2011multimodal}
J.~Ngiam, A.~Khosla, M.~Kim, J.~Nam, H.~Lee, and A.~Y. Ng.
\newblock Multimodal deep learning.
\newblock In {\em Proceedings of the 28th international conference on machine
  learning (ICML-11)}, pages 689--696, 2011.

\bibitem{norouzi2013zero}
M.~Norouzi, T.~Mikolov, S.~Bengio, Y.~Singer, J.~Shlens, A.~Frome, G.~S.
  Corrado, and J.~Dean.
\newblock Zero-shot learning by convex combination of semantic embeddings.
\newblock {\em arXiv preprint arXiv:1312.5650}, 2013.

\bibitem{pennington2014glove}
J.~Pennington, R.~Socher, and C.~D. Manning.
\newblock Glove: Global vectors for word representation.
\newblock In {\em EMNLP}, volume~14, pages 1532--1543, 2014.

\bibitem{razavian2014cnn}
A.~Razavian, H.~Azizpour, J.~Sullivan, and S.~Carlsson.
\newblock Cnn features off-the-shelf: an astounding baseline for recognition.
\newblock In {\em Proceedings of the IEEE Conference on Computer Vision and
  Pattern Recognition Workshops}, pages 806--813, 2014.

\bibitem{simonyan2014very}
K.~Simonyan and A.~Zisserman.
\newblock Very deep convolutional networks for large-scale image recognition.
\newblock {\em arXiv preprint arXiv:1409.1556}, 2014.

\bibitem{zhou2014learning}
B.~Zhou, A.~Lapedriza, J.~Xiao, A.~Torralba, and A.~Oliva.
\newblock Learning deep features for scene recognition using places database.
\newblock In {\em Advances in neural information processing systems}, pages
  487--495, 2014.

\end{thebibliography}

\end{document}